\documentclass[11pt,twoside]{article}
\usepackage{newpasp}

\usepackage{epsf}

\index{Fraternali, F.}
\index{Oosterloo, T.}
\index{Sancisi, R.}
\index{van Moorsel, G.}

\begin{document}

\title{The HI halo of NGC~2403}
\author{Filippo Fraternali}
\affil{I.R.A. (C.N.R.) \& Astronomy Department, Bologna, It}
\author{Tom Oosterloo}
\affil{N.F.R.A., Dwingeloo, NL}
\author{Renzo Sancisi}
\affil{Astronomical Observatory, Bologna, It \& Kapteyn Inst., Groningen, NL}
\author{Gustaaf van Moorsel}
\affil{N.R.A.O., Socorro, NM, USA}

\begin{abstract}
Deep VLA HI observations of the nearby spiral galaxy NGC~2403 reveal the
presence of an extended HI halo that shows slower rotation and a general
inflow ($\sim$ 15-25 km s$^{-1}$) towards
the center of the galaxy.
\end{abstract}


\section{The HI halo}

The position-velocity diagram along the major axis of the Sc galaxy NGC~2403 shows the presence of gas 
with anomalous velocities (see Fig. 3 of the paper by Sancisi {\it et al.}, this conference). 
This is the signature of an extended, slower rotating HI halo surrounding the thin disk (Schaap {\it et al.}, 2000)

We have isolated the `anomalous gas' in NGC~2403 by fitting a Gaussian line profile to the thin disk and subtracting it from the data.

\begin{figure}[htbp]
\plotfiddle{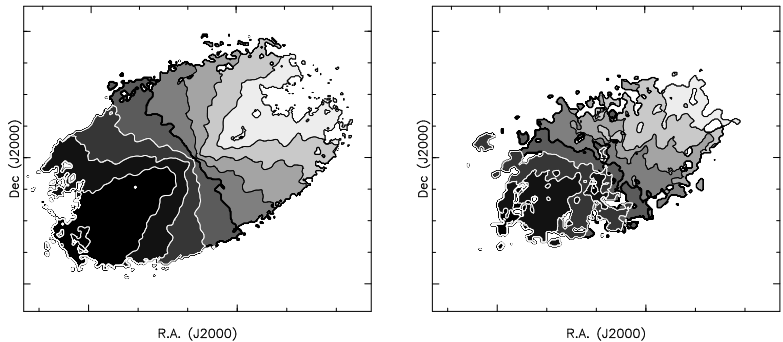}{4cm}{0}{70}{70}{-180}{-180}
\caption{\small Velocity fields of the total HI (left) and of the anomalous gas (right) for NGC~2403. The thick line shows the systemic velocity}
\label{fig1}
\end{figure}

Figure 1 shows the comparison between the velocity field of the total HI and that of the anomalous gas.
It is clear that the second has a major axis at a different position angle.
Moreover, the minor axis and the major axis are not orthogonal.
This may be the effect of a radial in-flow towards the centre of the galaxy.

\section{Models and discussion}

We have constructed various models of the HI layer of NGC~2403 adopting a two
component structure with a thin disk surrounded by a thicker,
slower rotating layer. In Figure 2 we show a position-velocity diagram
parallel to the minor axis at 2$'$ (South-West) from the center of the galaxy.
This is a suitable position to illustrate the effects of radial motions.

\begin{figure}[htbp]
\plotfiddle{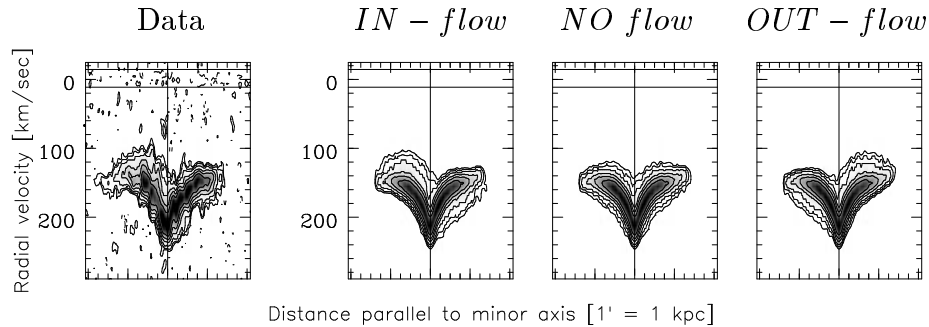}{4.5cm}{0}{110}{110}{-300}{-660}
\caption{\small Position-velocity diagram parallel to the minor axis of NGC~2403 at 2$'$ from the centre. Comparison between data and three models.}
\label{fig2}
\end{figure}

The diagram shows asymmetries especially visible at the low density
levels. which represent here the anomalous gas component. The models attempt
to reproduce such asymmetries by assuming radial motions.  The anomalous HI in
NGC~2403 clearly indicates a preference for an in-flow with a 
mean value of $\sim$20 km s$^{-1}$,
and somewhat larger values
in the central regions.

What is the origin of this gas? Possibilities are: 1) A galactic
fountain (Shapiro \& Field, 1976) in which the HI traces the
final phase of the fountain itself.  The large number of HI holes and
superbubbles in NGC~2403 is supporting evidence.  2) An accretion of {\it
primordial} extragalactic gas (Oort, 1970).  The kinematics observed in NGC~2403 
shows an analogy with some of the HVCs in our Galaxy.

\end{document}